\documentclass{PoS}

\newcommand{\be}{  \begin{eqnarray}}
\newcommand{\ee}{\end{eqnarray}  }

 \newcommand{\bmini}{\begin{minipage}}
 \newcommand{\emini}{\end{minipage}}

\newcommand\nn{\nonumber}

\title{Phase of the Fermion Determinant for QCD at Finite Chemical Potential}

\ShortTitle{Phase of Fermion Determinant}

\author{{K. Splittorff} \\
        Niels Bohr Institute, Blegdamsvej 17, Copenhagen, Denmark\\
        E-mail: \email{split@nbi.dk}}

\author{{J.J.M. Verbaarschot}\\
        Stony Brook University, Stony Brook, NY 11794\\
        E-mail: \email{jacobus.verbaarschot@stonybrook.edu}} 

\abstract{ In this lecture we discuss various properties of the phase
factor of the fermion determinant for QCD at nonzero chemical potential.
Its effect on physical observables is elucidated by
comparing the phase diagram of QCD and phase quenched QCD and by illustrating
the failure of the Banks-Casher formula with the example of one-dimensional
QCD. The average phase factor and the distribution of the phase are 
calculated to one-loop order in chiral perturbation
theory. In quantitative agreement with lattice QCD results, we find that 
the distribution is Gaussian with a width $\sim  \mu T \sqrt V$ (for
$m_\pi  \ll T \ll \Lambda_{\rm QCD}$).  
Finally, we introduce, so-called teflon plated observables which can be calculated
accurately by Monte Carlo even though the sign problem is severe.
}

\FullConference{The XXVI International Symposium on Lattice Field Theory\\
                 July 14-19 2008\\
                 Williamsburg, Virginia, USA}

\begin{document}

\section{Introduction}
Although the Euclidean formulation of QCD has been extremely successful
in the evaluation of observables from first principles,
at nonzero chemical potential, because of the phase of the fermion
determinant, progress has been slow and limited to a small part of the
QCD phase diagram (see for example \cite{owerev} for a recent review). 
Motivated by the drastic effect of the 
phase of fermion determinant on physical observables, we study its
properties to one-loop order in chiral perturbation theory. Since the
average phase factor is a ratio of two partition functions it has to vanish
exponentially 
with the volume when evaluated with respect to a positive definite
measure. Therefore, a sensible strategy is to extract observables by
extrapolation from small volumes. In the domain of validity of chiral
perturbation theory it possible to derive the volume dependence of
the average phase factor, and to determine the parameter range
where reweighting methods \cite{glasgow,fodor}
work. The distribution of the phase,
 which is also calculated, might be relevant in applications
of the density of states method \cite{azcoiti,ambj,schmidtf}.
It is less clear how methods that rely on analytical
continuation in the chemical potential \cite{Allton,gupta,lombardo,owe}
or methods that are based
on the canonical partition function \cite{kl,ejiric}
are affected by the sign problem,
but each method has its own difficulties.

With regards to the volume dependence of the QCD partition function
at $\mu \ne 0$, this can be investigated in detail in the microscopic
domain of QCD \cite{V,SV} where QCD is equivalent to chiral random
matrix theory. In this domain, the
cancellations that give rise to the discontinuity \cite{OSV} of the mass
dependence of the chiral condensate also take place for finite size matrices 
\cite{OSV2}.

After the definition of the average phase factor in section 2, 
we discuss its
effects on the phase diagram and on the chiral condensate in 1d QCD
in Section 3.
Results obtained with chiral perturbation theory including the distribution
of the phase are given in sections 4 and 5.

\section{Average Phase Factor}

At nonzero chemical potential, $\mu$, the QCD Dirac operator is given by
$D+\mu\gamma_0 +m$,
where $D$ is the anti-hermitian Dirac operator at zero chemical potential,
and $m$ is the quark mass. Because \\[0.1cm]
\bmini{8cm}
\includegraphics[width=3.4cm]{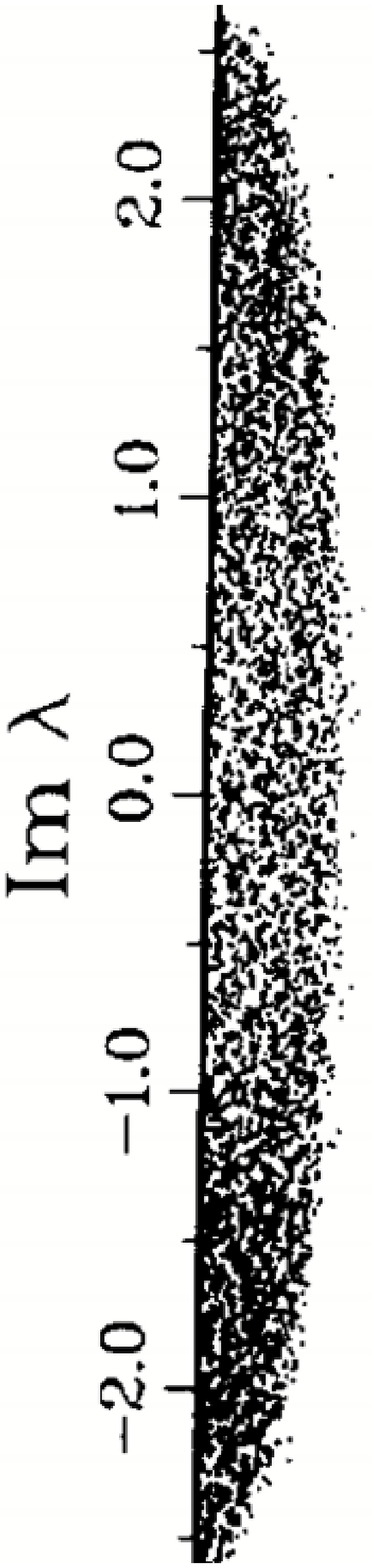}
\includegraphics[width=4.0cm]{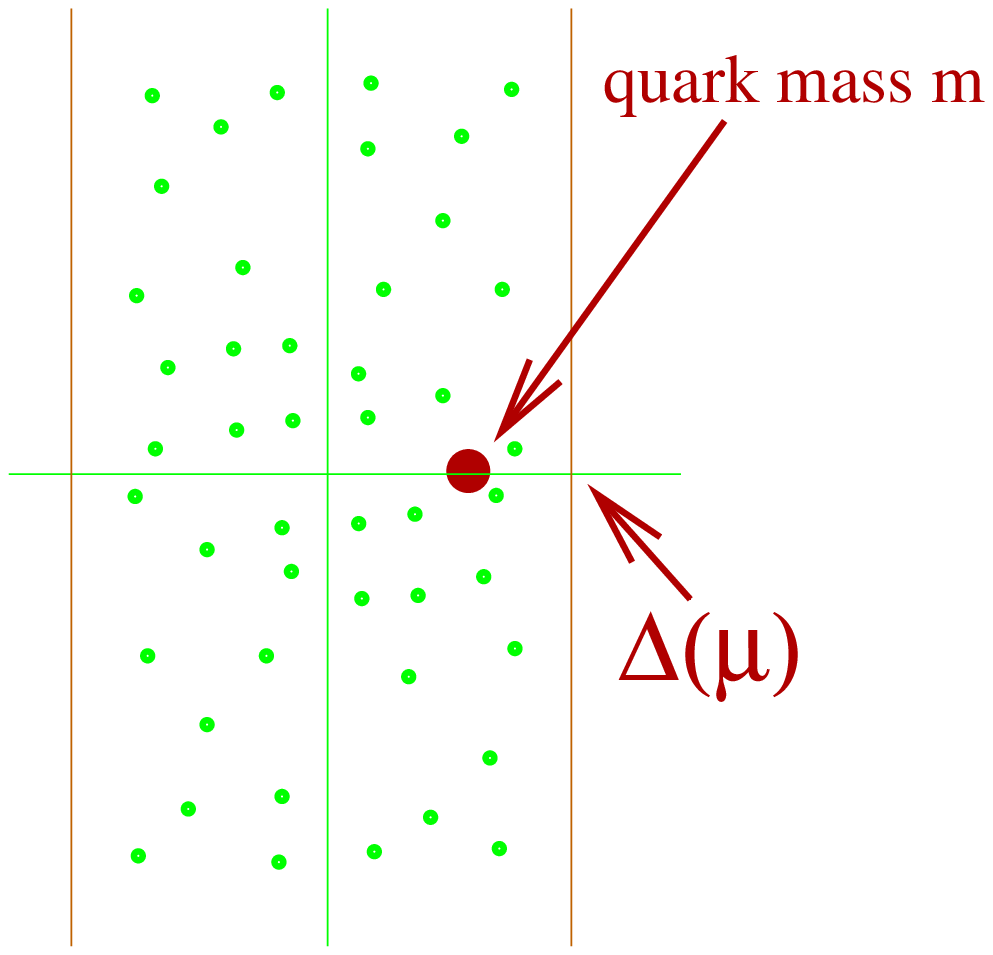}
\linespread{0.707}
\selectfont
{\fontsize{8}{8pt} \selectfont 
Fig. 1. Quenched Dirac eigenvalues on a $ 4^3\times 8 $ 
lattice \cite{all} (left)\\ and a schematic Dirac spectrum (right).}
\emini
\bmini{7cm}
 $\mu \gamma_0$ is hermitian, this Dirac operator does not have
any hermiticity properties and its eigenvalues are scattered in the complex
plane (See Fig. 1). 
Although the Dirac eigenvalues are still paired, the 
determinant is complex\\
$  
\det (D+\mu\gamma_0 +m) = e^{i\theta}|\det ( D+\mu\gamma_0 +m)|.
$\\
The phase will be strongly fluctuating if the quark
mass is inside the domain of eigenvalues resulting in an average phase
factor that vanishes in the thermodynamic limit. 
\emini\\[0.2cm]
The average can be
evaluated with respect to different partition functions. A particularly
useful definition of the average is with respect to the phase quenched
partition function,
\be
\langle e^{2i\theta}\rangle = 
\frac{\langle (\det(D+\mu\gamma_0 +m))^2 \rangle}
     {\langle |\det(D+\mu\gamma_0 +m)|^2 \rangle}=
\frac{Z^{\rm QCD}_{N_f=2}}{Z^{|{\rm QCD}|}_{N_f=2}},
\label{avPHfac}
\ee
which is the ratio of the QCD partition function and the phase quenched
QCD partition function. Because free energies are extensive and the free
energy of QCD and phase quenched QCD is generally different, the average
phase factor vanishes exponentially with the volume. This explains the
severity of the sign problem.

In addition to the sign problem, there is an overlap problem. For any given 
field configuration, the chiral condensate is close to the result
for the quenched or phase quenched partition function. The true value of the
condensate is due to exceptional gauge field configurations. The deep reason
for this is that determinants and the average phase factor depend exponentially
on the volume.

\section{The effect of the phase factor}

The simplest way to analyze the effect of the phase factor is to compare
the phase diagram of QCD and phase quenched QCD (denoted by |QCD|). 
First principle calculations for QCD
have only been done at or near the temperature axis of the phase diagram
in the $\mu T$-plane.
At zero temperature,  a transition to a phase
with nonzero baryon density takes place
at  $\mu \approx m_N/3$. Experimentally \cite{exp}, it has been shown
that this transition is first order with the first order line ending
in a critical endpoint (see Fig. 2). The remainder of the phase
diagram is based on general arguments and model studies. At extremely
high densities, one expects a superconducting phase whereas at intermediate
densities a confined phase with restored chiral symmetry is a 
likely possibility \cite{Jackson,McLerran1}.

\begin{center}
\includegraphics[width=10.0cm]{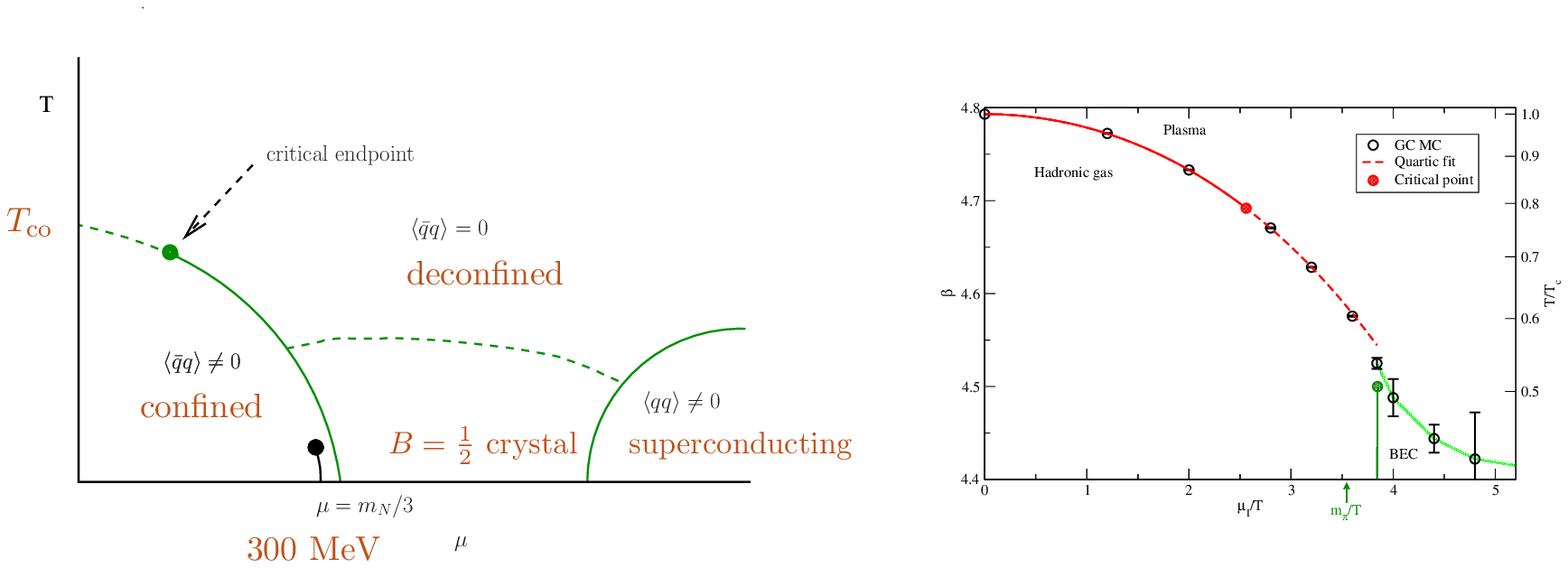} 
\end{center}
\linespread{0.707}
{\fontsize{8}{8pt} \selectfont 
Fig.~2.~The phase diagram of QCD in the temperature baryon chemical potential
plane (left) and in the temperature isospin chemical potential plane 
(right). The data points in the right figure are from lattice QCD simulations
\cite{wenger}.}

\vspace{2mm}

On inspection of the lattice Dirac spectra at nonzero chemical potential 
(see Fig. 1) we
notice that the eigenvalues are distributed more or less homogeneously
inside a strip and that the strip has a sharp edge. This implies that 
we can expect a phase transition when the quark mass hits the edge
of this strip located at  $\Delta(\mu)$. The critical chemical potential
is therefore given by $\Delta(\mu_c) = m$ \cite{Gibbs,TV}. 
Because phase quenched QCD is
QCD at nonzero isospin chemical potential, we know that at low temperatures
a phase transition to a pion condensed phase occurs at $\mu = m_\pi/2$. Using
the Gell-Mann-Oakes-Renner relation this results in a half-width of
the strip  of $\Delta(\mu) = 2\mu^2F^2/\Sigma$.

The width Dirac spectrum for each gauge field configuration in full QCD is also
given by $2\Delta(\mu)$, but instead of  a phase transition at 
$\mu = m_\pi/2$ there is a phase transition at $\mu = m_N/3$. In particular,
this implies that the chiral condensate should be a smooth function of
the chemical potential in the region with zero baryon density, and in the
low temperature limit, it should not depend at all on the chemical potential.
How this can happen in spite of the fact that the Dirac spectrum is
$\mu$-dependent has been understood in detail for the microscopic domain of
QCD \cite{OSV,OSV2}. Here we will illustrate this phenomenon for QCD in one-dimension
\cite{ravagli}.

\subsection{ Effect of the Phase in 1d QCD}

The Banks-Casher relation \cite{BC} states that
\be
\Sigma = \lim_{m\to 0} \lim_{V\to \infty}\frac {\pi \rho(m)}V,
\ee
where $\rho(\lambda)$ is the density of Dirac eigenvalues on the imaginary
axis and $V$ is the volume of space-time. Although originally intended for an
anti-Hermitian Dirac
operator, this relation correctly gives a vanishing chiral condensate 
for phase quenched QCD at $\mu \ne 0$ since there is no accumulation of
eigenvalues on the imaginary axis.
However, for full QCD at $ \mu \ne 0$, the chiral condensate has
a discontinuity when the quark mass crosses the imaginary axis, but 
it does so without an accumulation of eigenvalues on the
imaginary axis. The
alternative mechanism that is at work was first discovered
 in random matrix theory \cite{OSV,OSV2}. 
Below we will illustrate this mechanism for QCD in one dimension
which can also be viewed as a random matrix model. 

\centerline{\includegraphics[width=2.2cm]{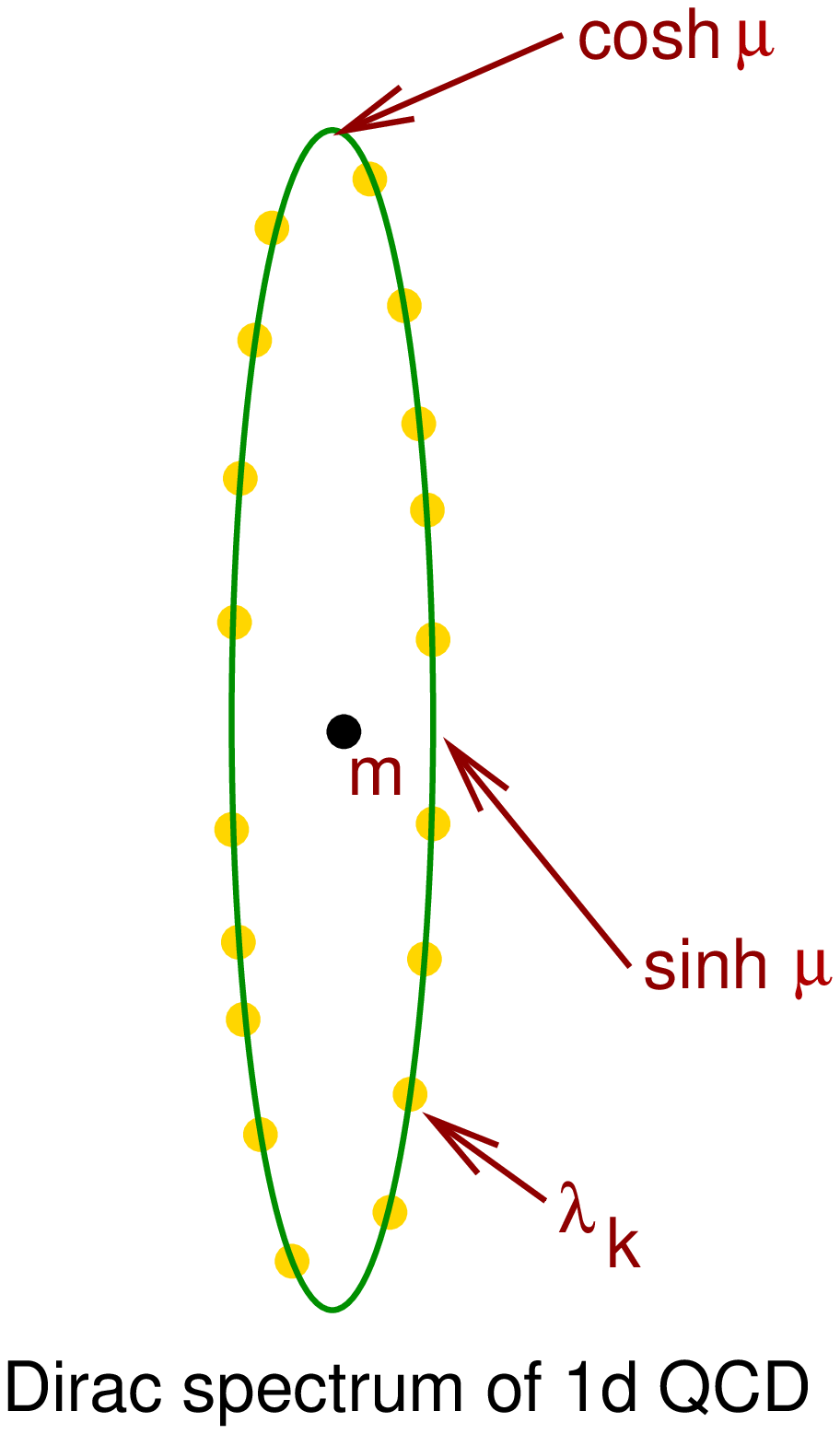}\hspace*{3cm}
\includegraphics[width=3.0cm]{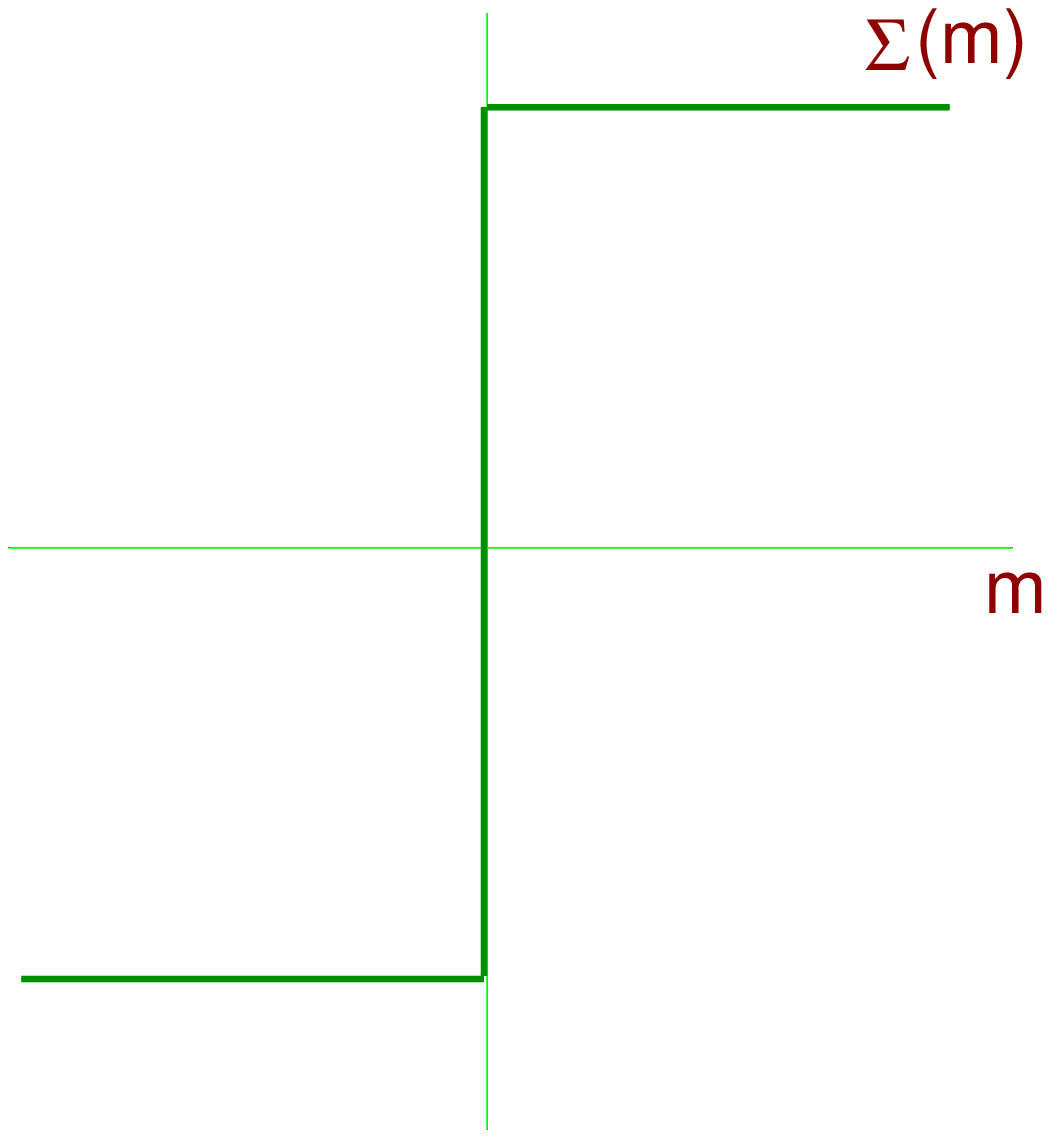}}
\linespread{0.707}
\selectfont
{\fontsize{8}{8pt} \selectfont
\noindent        Fig. 3.
Schematic plot of eigenvalues of the Dirac operator for 
lattice QCD in one dimension (left). The yellow dots denote the position of
the eigenvalues for a single gauge field configuration, whereas the green
ellipse shows the support of the spectrum in the thermodynamic limit.
The chiral condensate for one flavor versus the quark  mass is 
shown in the right figure. 
}

\linespread{1.0}\selectfont
The partition function of lattice QCD in one dimension is given by
\be
Z= \int_{U \in U(N_c)} dU \det D,
\ee
where the integral is over the Haar-measure of $U(N_c)$. The 1d lattice
Dirac operator is the $N\times N_c$ matrix
with hopping matrix elements given
by $U \exp(\mu)$ and its inverse \cite{BD} (the number of lattice points is
denoted by $N$). 
The chiral condensate for one flavor is
given by
\be\Sigma(m)  
=\frac {\left \langle \frac 1N\sum_k \frac 1{\lambda_k+m} 
{\prod_k(\lambda_k+m)}\right \rangle }
 {\left \langle
{\prod_k(\lambda_k+m)}\right \rangle }.
\ee
Since the eigenvalues, $\lambda_k$, are located on an ellipse in the complex
plane (see Fig. 3) the determinant has a complex phase. For $U(1)$ the chiral 
condensate can be evaluated analytically \cite{ravagli} with the result
that is shown in the right figure of Fig.~3. The amazing phenomenon,
also known as the ``Silver Blaze Problem'' \cite{cohen},
is that the chiral condensate is continuous when $m$ crosses the ellipse
of eigenvalues, but shows a discontinuity at $m =0$, where there are no
eigenvalues. This can happen because the chiral condensate is determined
by exponentially large (in the number of lattice points) contributions which
cancel to give a finite result for
$N \to \infty$. 

\section{Average Phase Factor in Chiral Perturbation Theory}

Because the average phase factor is the ratio of the QCD partition function
and the phase quenched QCD partition function, it can be evaluated by means of
chiral perturbation theory. The low-temperature limit of QCD is given by a
gas of pions and does not depend on $\mu$ to one-loop order.
Since the phase quenched partition function is QCD at non zero isospin
chemical potential the pions are charged with
respect to $\mu$ resulting in a $\mu$-dependent one-loop
result. For $N_f=2$  there
\bmini{6.5cm}
are two charged Goldstone modes $\pi_\pm$.
Denoting their one-loop contribution by
\be
 \prod_p \frac 1{\sqrt{ m_\pi^2+\vec p^2+(p_0-2i\mu)^2}} \equiv
e^{G_0(\mu)/2},\qquad
\ee
we obtain (the product is over all pions)
\be
\langle {\det}^2(D+m +\mu\gamma_0) \rangle &\sim&   
e^{-V F^{(0)}} \prod_k
e^{G_{0}(\mu=0)/2},\nn\\
\langle |\det (D+m +\mu\gamma_0)|^2 \rangle 
&\sim& e^{-V F^{(0)}} \prod_ke^{G_{0}(\mu)/2} .  \nn
 \ee
The average phase factor is the ratio of the two
partition functions cf.~(\ref{avPHfac}) resulting in the
cancellation of the neutral Goldstone bosons so that
\be
\langle e^{2i\theta} \rangle_{\rm pq}
&=& e^{G_0(\mu=0) - G_0(\mu)}.
\ee
\emini\hspace*{0.5cm}
\bmini{8.0cm}
\centerline{\includegraphics[width=7cm]{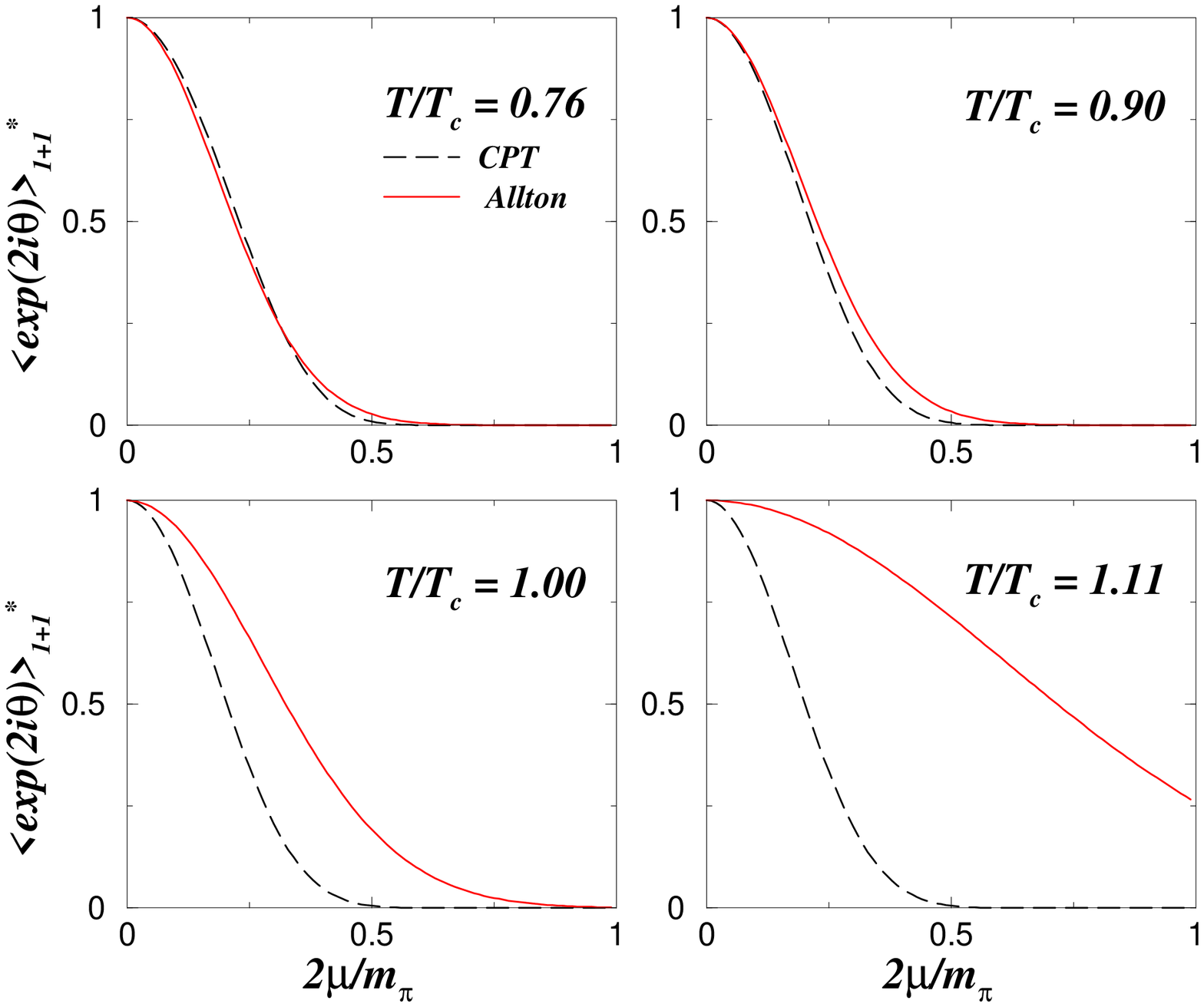}}
{\small  Fig. 4. Average phase factor in lattice QCD obtained from \cite{Allton} (red curve)
compared to one loop chiral perturbation
theory in a box equal to the size of the lattice (dashed curve) .}
\emini\\
In Fig. 4  where we compare one loop results for the average phase
factor to lattice QCD results obtained by Taylor expansion in the
chemical potential (there are no fitting parameters). The agreement
is striking even for $T/T_c=0.90$. As expected chiral perturbation theory fails
above $T_c$.

\section{Probability Distribution of the Phase}
The density of the phase angle is defined by 
$
\rho(\phi) = \langle \delta(\phi-
\small  {\rm Im}\log\det(D+m+\mu\gamma_0))
\rangle_{N_f}.
$
Notice that $ \phi \in \langle -\infty, \infty \rangle$.
If the average is over dynamical quarks, the phase density factorizes
into an overall phase factor and the phase density of the phase quenched
partition function ($\rho_{\rm pq}(\phi)$),
\be 
\langle \delta(\phi-\theta) e^{iN_f\theta}
|{\det}^{N_f}(D+m+\mu\gamma_0)|
\rangle 
=e^{iN_f\phi}
 \langle \delta(\phi-\theta) 
|{\det}^{N_f}(D+m+\mu\gamma_0)|
\rangle .
\ee
According to the Central Limit Theorem we 
expect that $\rho_{\rm pq}(\phi)$ approaches a Gaussian distribution \cite{Ejiri}. 
Notice however that observables are determined by correlations 
with the phase of the fermion
determinant, and  knowing the Gaussian distribution is only part of the story.
 
The phase density can be evaluated using the replica trick and chiral 
perturbation theory \cite{splitphase}. Fourier 
transforming the $\delta$-function we obtain
\be
\rho_{N_f}(\phi) &=& \langle \delta(\phi-{\rm Im}\log\det(D+m+\mu\gamma_0)) 
\rangle_{N_f} 
= \langle \sum_n e^{in(\phi-{\rm  Im}\log\det(D+m+\mu\gamma_0)} \rangle_{N_f}
 \ee
The phase density therefore follows from the moments of the phase factor,
\be
\langle e^{2in\theta}\rangle_{N_f} =\frac 1{Z_{N_f}}
\left\langle \frac{{\det}^{n+N_f}(D+m+\mu\gamma_0)} 
{{\det}^n(D^\dagger+m+\mu\gamma_0)}\right \rangle ,
\ee
which is the ratio of two partition functions
with a numerator that contains 
both bosonic and fermionic Goldstone particles. All
$2n(n+N_f)$ charged Goldstone particles are fermions while
the uncharged Goldstone particles are bosons. We thus find
\be
\langle e^{2in\theta}\rangle_{N_f} =e^{n(n+N_f)[G_0(\mu=0) -G_0(\mu)]}.
\ee
By Poisson resummation we obtain ($\Delta G=G_0(\mu)-G_0(\mu=0)$)\\
\bmini{8.0cm}
\be
\rho(\phi)=\sum_n e^{in\phi} e^{-n/2(n/2+N_f) \Delta G}
=\frac {e^{ \frac 14 N_f^2 \Delta G }}{\sqrt{\pi\Delta G} }\,e^{iN_f\phi 
- \frac {\phi^2}{\Delta G}}.\;\;\;\quad
\ee 
A simple expression for the width of the Gaussian is obtained in a small
$\mu$ expansion of the ChPT result. 
For the extensive quantity $\Delta G$ we find ($m_\pi\ll T$) 
\be
\Delta G \sim V \mu^2 T^2.
\ee 
Since the width of the Gaussian $\sim \sqrt{\Delta G}$ this result
is in quantitative agreement with the lattice results by Ejiri
\cite{Ejiri} (See Fig. 5).
\emini\hspace*{0.5cm}
\bmini{6.5cm}
\begin{center}
\includegraphics[width=6cm]{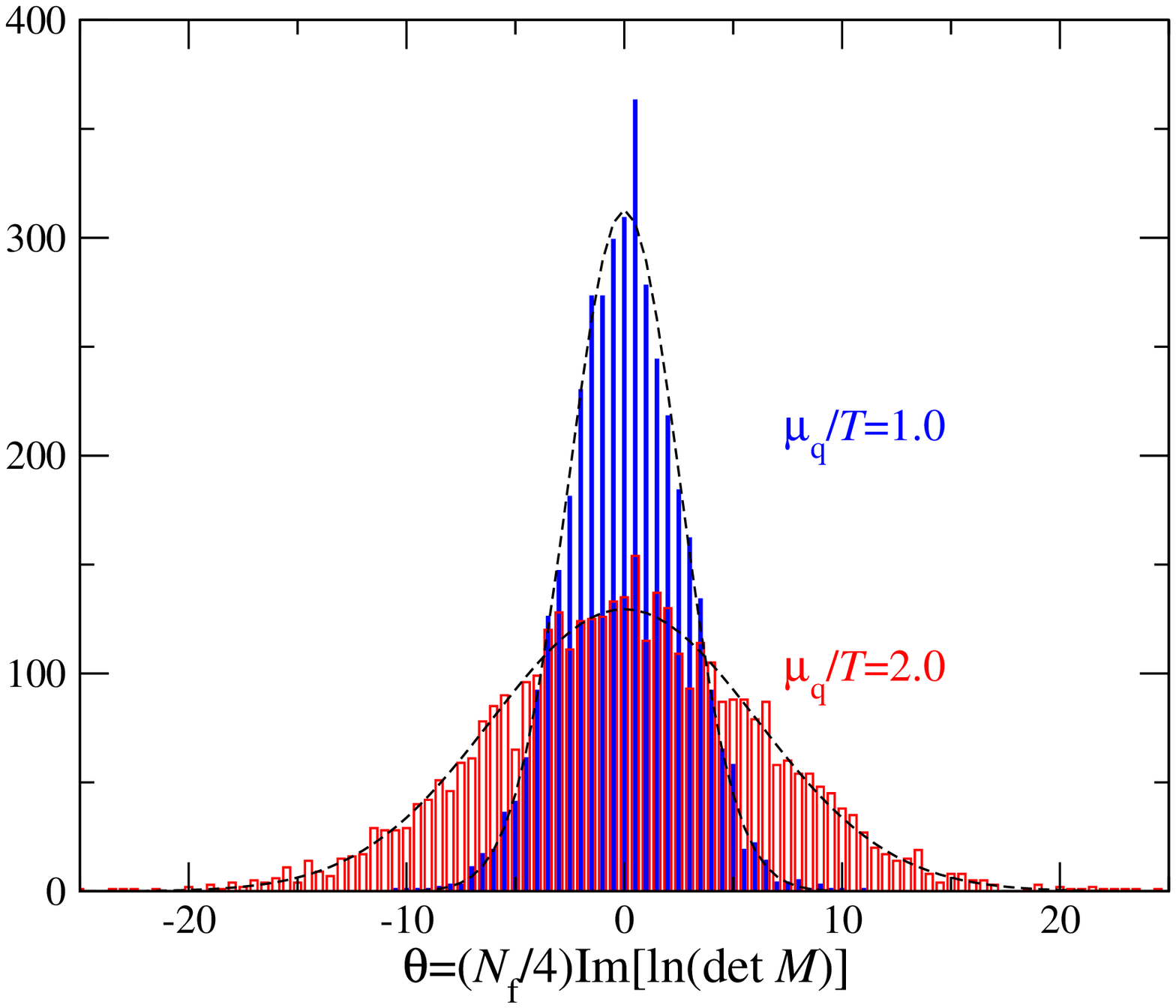}
\end{center}\vspace*{-0.5cm}
{ \small Fig. 5. Phase density in lattice QCD \cite{Ejiri}.}
\emini\\
The average phase factor can also be evaluated in a weak coupling
expansion valid at high temperatures \cite{vuorinen,rebhan}. The 
small-$\mu$ limit of the first nonvanishing contribution is given
by \cite{splitphase}
\be
F_{\rm QCD} (\mu,T)-F_{\rm |QCD|} (\mu,T)\sim V \alpha_s^2 \mu^2 T^2,
\ee
with $V$ the space-time volume. This implies that the sign problem remains
severe in the thermodynamic limit of the weak coupling result. Notice that
expectation values can be calculated accurately despite the sign problem if
the correlations between the observable and the phase factor are small. 
Such {\sl teflon plated observables} should obey the condition (here $N_f=2$)
\be
\frac{|\langle {\cal O} e^{2i\theta} \rangle_{|{\rm QCD}|} -
\langle {\cal O} \rangle_{|{\rm QCD}|}
\langle e^{2i\theta} \rangle_{|{\rm QCD}|}|}
{\langle {\cal O}\rangle_{|{\rm QCD}|}
\langle e^{2i\theta}\rangle_{|{\rm QCD}|} }
\ll 1 
\ee
This condition only makes sense if 
$\langle {\cal O} \rangle_{\rm{|QCD|}} \ne 0$ as can be seen by
rewriting the l.h.s. as
\be
\frac{|\langle {\cal O} \rangle_{\rm{QCD}}
-\langle {\cal O} \rangle_{\rm{|QCD|} }|}
{\langle {\cal O} \rangle_{\rm{|QCD|}}}.
\ee
For example, when ${\cal O}$ is the baryon density this ratio is 
$O(\alpha_s^2)$ whereas $\langle {\cal O} \rangle_{\rm{QCD}}$
 is  $O(\alpha_s^0)$
\cite{maria}.

\section{Conclusions}
The phase factor of the fermion determinant completely alters the 
properties and the physics of the QCD partition function at nonzero
chemical potential. We have illustrated this by the relation between
the chiral condensate and the Dirac spectrum and the difference between
the phase diagram of QCD and phase quenched QCD. Using one-loop chiral
perturbation theory, we have  determined both the magnitude of the
average phase factor and the statistical distribution of the phase
which turns out be a Gaussian with width $\sim \mu T \sqrt V$. Finally,
we introduced a class of observables (teflon plated observables)
that can be evaluated accurately despite the sign problem. An example
of such observable is the baryon density at high temperature and small
chemical potential.

\noindent{\bf Acknowledgments:} 
This work was
supported in part by U.S. DOE Grant No. DE-FAG-88ER40388.

\end{document}